	\documentclass[preprint]{seismica}

	\usepackage{amsmath,amssymb,bm}

	\newcommand{\ten}[1]{$10^{#1}$}

	\newcommand{\mo}[1]{\operatorname{{#1}}}
	\renewcommand{\d}{$\bm{d}$}
	\newcommand{\m}{$\bm{m}$}
	
	\newcommand{\design}{$ \bm{\xi}$}

	\newcommand{\eg}{e.\,g.,\,}

	\title{Near-real-time design of experiments for seismic monitoring of volcanoes}

	\author[1]{Dominik Strutz
		\thanks{Corresponding author: dstrutz@ed.ac.uk}
	}
	\author[1]{Andrew Curtis
	}
	\affil[1]{School of GeoSciences, University of Edinburgh, Edinburgh, UK}

\begin{document}

		\makeseistitle{
			\begin{summary}{Abstract}\label{sec:abstract}
				Monitoring the seismic activity of volcanoes is crucial for hazard assessment and eruption forecasting. The layout of each seismic network determines the information content of recorded data about volcanic earthquakes, and experimental design methods optimise sensor locations to maximise that information. We provide a code package that implements Bayesian experimental design to optimise seismometer networks to locate seismicity at any volcano, and a practical guide to make this easily and rapidly implementable by any volcano seismologist. This work is the first to optimise travel-time, amplitude and array source location methods simultaneously, making it suitable for a wide range of volcano monitoring scenarios. The code-package is designed to be straightforward to use and can be adapted to a wide range of scenarios, and automatically links to existing global databases of topography and properties of volcanoes worldwide to allow rapid deployment. Any user should be able to obtain an initial design within minutes using a combination of generic and volcano-specific information to guide the design process, and to refine the design for their specific scenario within hours, if more specific prior information is available.
			\end{summary}
		} 

		\section{Introduction}\label{sec:introduction}
			Forecasting episodes of volcanic unrest is a significant challenge due to the highly variable characteristics of volcanic systems \citep{Chiodini2016-dd, Selva2015-lp, Marzocchi2012-nw, Sigmundsson2010-jh}. Monitoring volcanoes serves two main purposes: first, to understand the structure and dynamics of each volcano, as this is crucial for hazard assessment, eruption forecasting, and early warning, and second, for risk mitigation at times of volcanic unrest \citep{Pallister2015-ty}. Monitoring allows accelerating rates of seismic activity to be detected, which often occurs before eruptions \citep{Sigmundsson2010-jh, Saltogianni2014-iy, Cannavo2015-um}, and the location of volcano-tectonic earthquakes to be tracked in near real-time, which may facilitate eruption forecasting \citep{Sigmundsson2015-xz,Pallister2015-ty, Falsaperla2015-zh}. In addition, long-period earthquakes and micro-earthquakes can be key indicators of imminent eruptions, especially during extended eruptive sequences \citep{McNutt2005-pr}. In all cases, the location of seismic sources is crucial to allow the underlying processes to be better understood.

			Seismicity is usually monitored by a sparse network of seismometers deployed around the volcano. The layout of this network (the experimental design) determines the information content of recorded data, and therefore the quality of derived or interpreted information. The goal of the experimental design process is to optimise the network layout, and sensor types to maximise the information that can be gained from the data \citep{Maurer2010-jt, Strutz2023-pl}. Despite the importance of experimental design, and the fact that optimal design methods have existed for decades \citep[e.g.,][]{Steinberg1995-ox}, in practice, seismic networks are rarely optimised taking the known physics and uncertainty of the problem into account. This is likely to be due to a lack of awareness of the potential for design methods to enhance survey results, a lack of expertise in understanding the range of design methods and their relative merits \citep{Bloem2020-gp}, or a lack of time in rapid deployment operations.
			
			The goal of this paper is to provide a practical guide to the implementation of Bayesian experimental design for volcano seismic monitoring, and a code package that makes this easily and rapidly implementable by any volcano seismologist. This work is the first to take travel-time, amplitude and array source location methods into account simultaneously, making it suitable for a wide range of volcano monitoring scenarios. The code package is implemented in Python, is available as open source software, and makes use of existing global databases of topography and properties of volcanic systems to provide designs rapidly. The software is designed to be straightforward to use and can easily be adapted to a wide range of scenarios. Any user should be able to obtain a first design within minutes using a combination of generic and volcano specific information to guide the design process, and be able to refine the design for their specific scenario within hours if they have more specific prior information available.

			In this paper we first introduce the fundamentals of Bayesian experimental design in the context of monitoring volcano induced seismicity in section \ref{sec:methods}. We present an example design process in section \ref{sec:example_design_process} where we show how to define the prior distribution, the physical model, how to calculate the expected information gain, and how to optimise the experimental design. We then discuss how to interpret the results, and some advanced topics such as the use of heterogenous velocity models in section \ref{sec:heterogenous_velocity_model}.

		\section{Methods}\label{sec:methods}
			This section introduces fundamental theory for Bayesian experimental design for locating and monitoring volcano-induced seismicity. The focus of this paper is to provide a practical guide to the implementation of design methods in real-world scenarios. To keep this section concise and suitable for a broad audience, we do not go into theoretical details, but we refer the reader to relevant literature for a more fundamental introduction.

			\subsection{Bayesian Experimental Design}\label{sec:bayesian_experimental_design}
			When designing an experiment we aim to find a set of observation points described by a vector \design{} (the design) that maximises the amount of information that we expect to gain by collecting data at these points. In the context of volcano seismic monitoring, the goal is usually to find a set of station/array locations that maximises the expected information about the location of seismic sources. To account for the full uncertainty and non-linearity of the problem, we use a Bayesian framework for the experimental design process.

				\subsubsection{Bayesian Inference}\label{sec:bayesian_inference}
					Before introducing the experimental design problem, we briefly introduce Bayesian inference. The goal of Bayesian inference is to infer the uncertainty in model parameters \m{} given some data \d{} observed using the experimental design \design{}. This uncertainty is characterised by a so-called posterior probability distribution $p(\bm{m} \, | \, \bm{d}, \bm{\xi})$. Bayes' theorem states that the posterior distribution can be calculated by
					\begin{equation}
						p(\mathbf{m} \, | \, \mathbf{d}, \mathbf{\xi}) = \frac{p(\mathbf{d} \, | \, \mathbf{m}, \mathbf{\xi}) \, p(\mathbf{m})}{p(\mathbf{d} \, | \, \mathbf{\xi})}
					\end{equation}
					where $p(\bm{d} \, | \, \bm{m}, \bm{\xi})${} is the likelihood of recording data \d{} given the particular set of parameter values defined by \m{}, $p(\bm{m})$ is the prior probability of \m{}, and $p(\bm{d} \, | \, \bm{\xi})$ is called the evidence of data \d{}. All terms that involve observed data also depend on design $\xi$.

					The likelihood is the probability of observing the recorded data given the model parameter values (i.e., assuming that those values are true) and the particular experimental design that was deployed. In our case, the likelihood is the probability of observing a set of travel-times, amplitudes, or array measurements given a source location and the experimental design. The likelihood is assumed to be a multivariate Gaussian, where the mean is the data predicted using the forward model, and the covariance matrix is diagonal with appropriate entries for expected measurement uncertainties given the data types and receiver/array locations in $\xi$.

					The prior distribution describes where we believe seismicity is most likely to occur in the subsurface before we have observed any data. To make the experimental design process as general as possible, the prior distribution in this study is defined on a discrete grid that spans a subset of the subsurface around each volcano. The distribution is assumed to be uniform within each grid cell. This description makes it straightforward to define the prior distribution, and to sample from it. In section \ref{sec:prior_information}, we show how this grid can be defined and refined using readily available information about any volcano of interest.

				\subsubsection{Bayesian Experimental Design}\label{sec:bayesian_experimental_design_subsubsection}
					The Bayesian framework makes it straightforward to estimate the expected increase in information to be gained by collecting data at a specific set of observation points. This expected information gain (EIG) can be expressed as the average of the information described by the posterior distribution $p(\bm{m} \, | \, \bm{d}, \bm{\xi})$ compared to the information in the prior distribution $p(\bm{m})$ \citep{Lindley1956-lu}:
					\begin{equation}
						\mo{EIG} = \mathbb{E}_{p(\bm{d} | \bm{\xi})} \Bigl\{ \mo{I}[p(\bm{m} \, | \, \bm{d},\bm{\xi})] - \mo{I}[p({\mathbf{m}})] \Bigr\} \label{eqn:EIG_def_model}
					\end{equation}
					where $\mo{I}[p]$ is the Shannon information \citep{Shannon1948-of} of distribution $p$ (for more information see appendix \ref{app:shannon_information}), and $ \mathbb{E}_{p(\bm{d} \, | \,\bm{\xi})}$ is the expectation over all possible data that is likely to be observed given the available prior information. Intuitively, equation \ref{eqn:EIG_def_model} states that the information gain is the reduction in uncertainty in source locations \m{} given observations \d{} at stations/arrays \design{}, compared to their uncertainty given only the prior information; since no observed data is usually available for the design process we average over all possible observations (or a representative subset) that are likely to be recorded. We obtain the subset of likely observations by forward modelling data for a set of earthquake locations sampled from the prior distribution, and adding simulated observational noise described by the data likelihood.					
					\begin{figure*}[ht!]
						\centering
						\includegraphics[width=\textwidth]{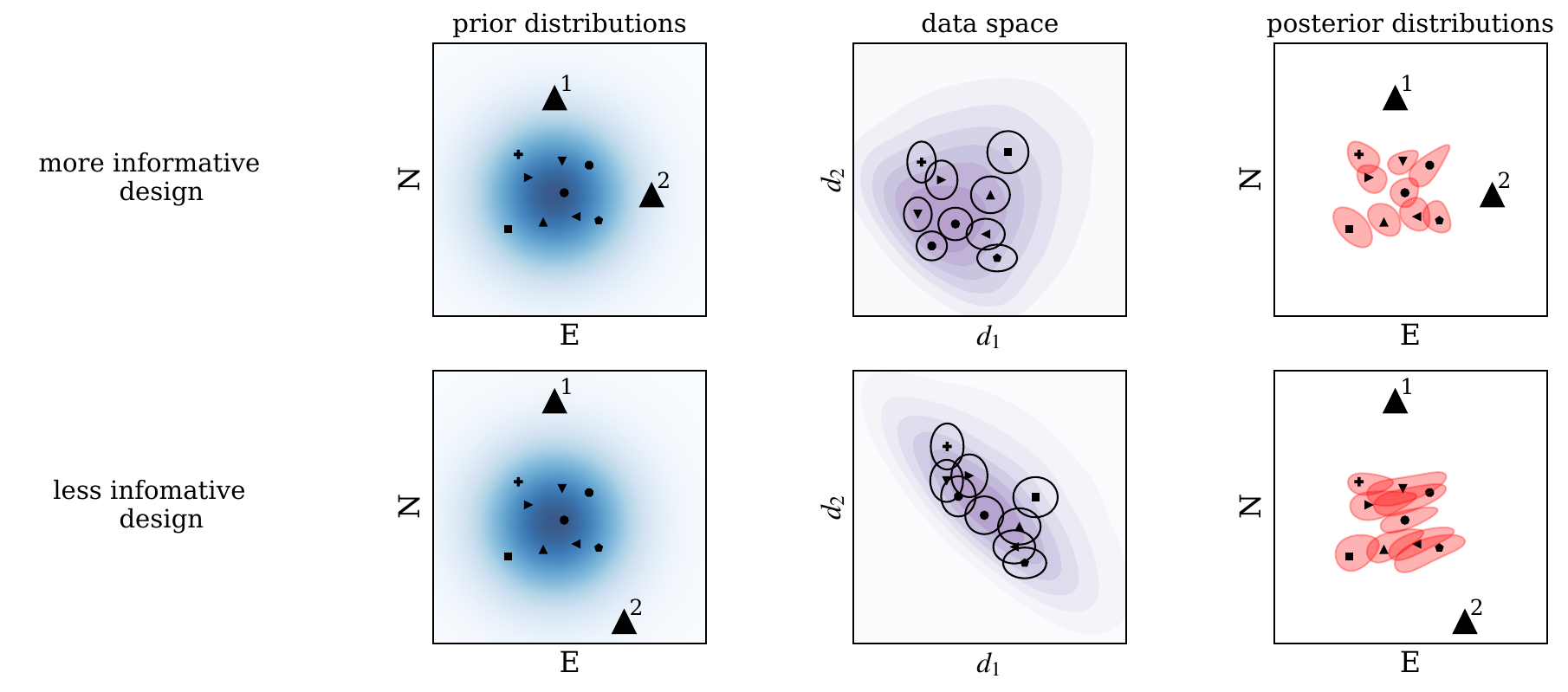}
						\caption{Illustration of the expected information gain (EIG) in the Bayesian experimental design process. The left panel shows a set of models sampled from the prior distribution; the middle panel shows the corresponding observed data and its likelihood (two standard deviation ellipses), and in the background the contours of the evidence calculated using a large number of data samples; and the right panel shows a contour of the posterior distribution for each model. Each model and data sample has a different symbol to allow corresponding pairs of models and data samples to be identified; in the left and middle panels deeper colours represent higher probabilities.}		
						\label{fig:EIG_intuitive_schematic}
					\end{figure*}

					Using the EIG as our design objective function, the best design can be expressed mathematically as
					\begin{align}
						\bm{\xi}^{*} &= \underset{\bm{\xi} \in \Xi}{\arg \max } \; \mo{EIG} (\bm{\xi}) \label{eqn:eig_optimization}
					\end{align}
					where $\Xi$ is the set of all possible experimental designs, and $\arg \max$ is mathematical shorthand for finding the design that maximises the EIG with respect to $\xi$. In this work, we use a genetic algorithm \citep{Holland1975-wf, Sambridge1993-sr, Gad2023-mq} to optimise the experimental design; this is a population-based optimisation algorithm that operates analagously to processes of natural selection amongst a set of potential designs. It is well suited for experimental design problems since it can optimise designs with mixed data types and constraints. The genetic algorithm is a stochastic (pseudo-random) algorithm, which means that the results can vary between runs. We use a fixed seed for the random number generator to ensure that the results are reproducible.

					To evaluate the EIG in the form of Equation \eqref{eqn:EIG_def_model}, we would need to calculate the posterior distribution $p(\bm{m} \, | \, \bm{d}, \bm{\xi})$ for each possible observation \d{}, which is computationally very expensive (unless we approximate the posterior distribution \citep{Foster2019-rx}). To make the problem tractable we rearrange the EIG to explicitly depend on the evidence $p(\bm{d} \, | \, \bm{\xi})$ and the likelihood $p(\bm{d} \, | \, \bm{m}, \bm{\xi})${} instead of the posterior distribution $p(\bm{m} \, | \, \bm{d}, \bm{\xi})$ and the prior distribution $p(\bm{m})$ (details in appendix \ref{app:rearange_EIG}). The EIG can then be expressed as:
					\begin{equation}
						\mo{EIG}(\xi) = \mathbb{E}_{p(\bm{m})} \Bigl\{ \operatorname{I}[p(\bm{d} \, | \, \bm{m},\bm{\xi})] \Bigr\}- \mo{I}[p(\bm{d} \, | \,\bm{\xi})] \label{eqn:EIG_def_data}
					\end{equation}
					where the expectation is now over the prior distribution $p(\bm{m})$.

					We can evaluate the EIG in equation \ref{eqn:EIG_def_data} with the nested Monte-Carlo (NMC) method \citep{Ryan2003-qp}. We use the computationally more efficient variation of \citet{Huan2013-nf} which is defined as follows:
					\begin{align}
						\begin{split}
						\mo{EIG}_{\mathrm{NMC}} = \frac{1}{N} \sum_{i=1}^{N} \Bigl\{ & \log \left[ p\left(\bm{d}_{i} \, | \, \bm{m}_{i}, \bm{\xi} \right) \right] - \log \left[ p\left(\bm{d}_{i} \, | \, \bm{\xi} \right) \right] \Bigr\} \\
						= \frac{1}{N} \sum_{i=1}^{N} \Biggl\{ & \log \left[ p\left(\bm{d}_{i} \, | \, \bm{m}_{i}, \bm{\xi} \right) \right] - \\ & \log \left[ \frac{1}{N} \sum_{j=1}^{N} p\left(\bm{d}_{i} \, | \, \bm{m}_{j}, \bm{\xi} \right) \right] \Biggr\} \label{eqn:NMC_reuse}
						\end{split}
					\end{align}
					where we sample $N$ models $\{\bm{m}_{i} \}_{i=1}^{N}$ from the prior distribution $p(\bm{m})$, calculate $N$ data vectors $\{\bm{d}_{i} \}_{i=1}^{N}$, one vector for each sampled model, and then approximat the EIG of Equation \eqref{eqn:EIG_def_data} by replacing the expectation with the average $\frac{1}{N} \sum_{i=1}^{N}$ over the $N$ samples.

					Calculating the EIG using the NMC method, while unbiased, is computationally expensive. During the design optimisation process we must evaluate this quantity many times, which can become intractable. We therefore invoke the $\mathrm{D}_N$ method \citep{Coles2011-ks} in which we approximate $p(\bm{d} \, | \, \bm{\xi})$ by a multivariate Gaussian distribution. This results in the following expression for the EIG:
					\begin{align}
						\begin{split}
						\mo{EIG}_{\mathrm{DN}} = \frac{1}{N} & \sum_{i=1}^{N} \log \left[ p\left(\bm{d}_{i} \, | \, \bm{m}_{i}, \bm{\xi} \right) \right] - \\ & \frac{3}{2} \left(1 + \log(2\pi) \right) + \frac{1}{2} \log \left| \bm{C}_{\bm{d}} \right| \label{eqn:DN}
					\end{split}
					\end{align}
					where we again sample $N$ models $\{\bm{m}_{i} \}_{i=1}^{N}$ from the prior distribution $p(\bm{m})$ and calculate corresponding data samples $\{\bm{d}_{i} \}_{i=1}^{N}$. Instead of a nested loop, we now calculate the information in the evidence by making use of the fact that the information in a multivariate Gaussian distribution is given by
					\begin{equation}
						\mo{I}[p\left(\bm{d}_{i} \, | \, \bm{\xi} \right)] = \frac{k}{2} \left(1 + \log(2\pi) \right) + \frac{1}{2} \log \left| \bm{C}_{\bm{d}} \right| \label{eqn:info_gaussian}
					\end{equation}
					where $k$ is the dimension of the data space and $\bm{C}_{\bm{d}}$ is the covariance matrix of the data samples $\{\bm{d}_{i} \}_{i=1}^{N}$. The $\mathrm{D}_N$ method is computationally far more efficient than the NMC method, but offers an upper bound for the true EIG and therefore overestimates the EIG. Despite this bias, the method has been shown to perform well in the design of seismic monitoring networks since only relative values of the EIG for different designs are important \citep{Bloem2020-gp}. Thus, unless otherwise stated, in this work we use the $\mathrm{D}_N$ method to calculate the EIG during the design optimisation process, and use the NMC method to validate and interpret the results.

					To provide a more intuitive understanding of the EIG, especially of its formulation in data space, we consider a schematic example, in which we demonstrate how more or less informative designs affect terms in Equation \eqref{eqn:EIG_def_data}. The first step is to sample a set of model samples from the prior distribution $p(\bm{m})$, shown in the left panel of Figure \ref{fig:EIG_intuitive_schematic}. Then we calculate the data samples (including a realisation of measurement noise described by the likelihood) that would be observed for each model: here, the distance between model and receiver locations, which is equivalent (proportional) to a travel-time from each source to receivers in a homogenous medium, as shown in the middle panel of Figure \ref{fig:EIG_intuitive_schematic}. The density of the data sample realisations allows the evidence (calculated using a large number of data samples in the background and shown as purple contours in the middle panel of Figure \ref{fig:EIG_intuitive_schematic}) to be visualised. Each data point has an associated likelihood, displayed as the two standard deviation ellipse. This shows graphically how different designs affect the evidence, the likelihood, and, therefore, the EIG.

					The term 
					\begin{align}
						\mathbb{E}_{p(\bm{m})} \operatorname{I}[p(\bm{d} \, | \, \bm{m},\bm{\xi})] = \frac{1}{N} \sum_{i=1}^{N} \log \left\{ p\left(\bm{d}_{i} \, | \, \bm{m}_{i}, \bm{\xi} \right) \right\} \label{eqn:expected_likelihood}
					\end{align}
					computes the expected information content in the data likelihood by averaging the probability of observing the data samples given the (true) source location model samples. This information content will be maximised for small standard deviations and minimised for large standard deviations (small and large ellipses, respectively, in the middle panel of Figure \ref{fig:EIG_intuitive_schematic}). Since we maximise this quantity, the standard deviation of the data samples should be as small as possible.
				
					The evidence term 
					\begin{align}
						\operatorname{I}[p(\bm{d} \, | \, \bm{\xi})] = \frac{1}{N} \sum_{i=1}^{N} \log\left[ \frac{1}{N} \sum_{j=1}^{N} p\left(\bm{d}_{i} \, | \, \bm{m}_{j}, \bm{\xi} \right) \right] \label{eqn:evidence_nmc}
					\end{align}
					or its approximation 
					\begin{align}
						\operatorname{I}[p(\bm{d} \, | \, \bm{\xi})] \approx \frac{3}{2} \left(1 + \log(2\pi) \right) + \frac{1}{2} \log \left| \bm{C}_{\bm{d}} \right| \label{eqn:evidence_gaussian}
					\end{align}
					calculates the information content in the evidence, which here is proportional to the inverse of the spread of the data samples (indicated by contours calulated using a large number of samples in the background). A large spread is preferable since it indicates that the data samples are well spread out and can be more easily distinguished from each other (or more formally, that the information in the evidence is low).
					
					The EIG is the difference between the result of evaluating equation \ref{eqn:expected_likelihood} and equation \ref{eqn:evidence_nmc} or its approximation in equation \ref{eqn:evidence_gaussian}, which is the difference between the average likelihood of the data samples and the measure of the inverse of the spread of the data samples. If data samples are closer together, this means that for fixed data values measured by an experiment, it is hard to distinguish between the source location model that generated those data, and other source location samples - which leads to a low EIG; intuitively this would indicate that the posterior distribution is less spread-out, as shown in the bottom right panel of Figure \ref{fig:EIG_intuitive_schematic}. On the other hand, if a different experimental design caused the data samples to be more spread out such that their uncertainties do not overlap, the EIG will be high and each source location can be discriminated from the others given a measured data set. This is clearly illustrated in the top right panel of Figure \ref{fig:EIG_intuitive_schematic}.

			\subsection{Seismic Source Location Methods}\label{sec:seismic_source_location_methods}
				Volcanoes are complex systems driven by a wide variety of magmatic and hydrothermal processes, and this leads to a wide range of characters of seismic activity that can be detected at the surface \citep{Zobin2017-pr}. Locating these seismic sources is a crucial step in understanding the underlying processes. For the observed signals, the characteristic periods span \ten{-1} to \ten{2}~s and corresponding wavelengths span \ten{2} to \ten{5}~m \citep{Saccorotti2021-cu}. This wide range of scales leads to a variety of methods for locating seismic sources.
				\begin{figure}[ht!]
					\includegraphics[width=8.6cm]{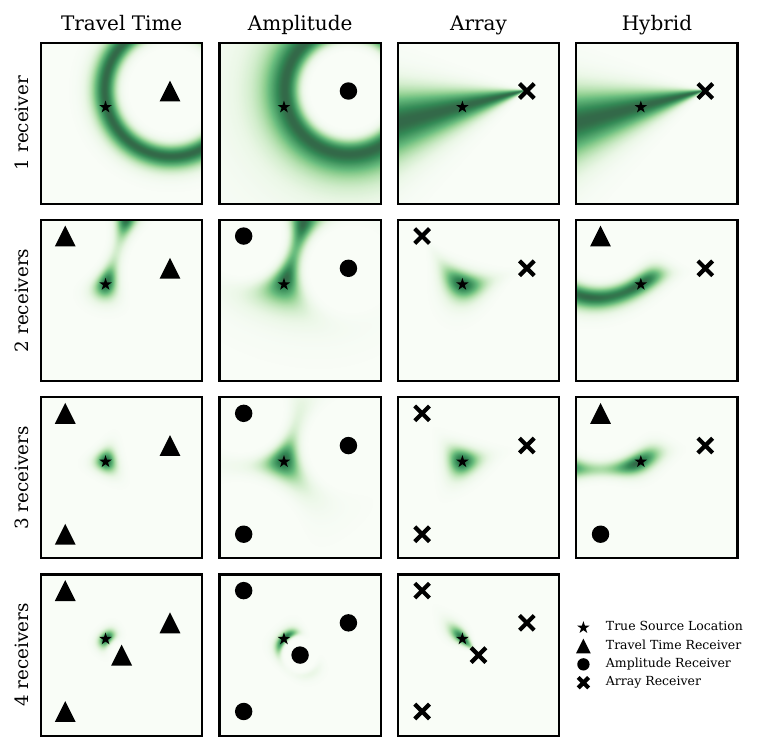}
					\caption{Example of likelihoods (dark green indicates high likelihood) for travel-time data from a synthetic true event for up to four stations at the same locations for different data types. For details on the different data types see section \ref{sec:seismic_source_location_methods}. The hybrid column shows how the likelihoods of different data types (represented by symbols) can be combined. In row 1 the hybrid and array results are identical since only a single measurement type is available.}
					\label{fig:likelihood_examples}
				\end{figure}

				This paper focuses exclusively on data derived from waveforms rather than on waveform data themselves, since the optimal design process and the forward model for full waveforms are computationally costly. Additionally, waveform-based methods are often very sensitive to the subsurface velocity model \citep[e.g.][]{O-Brien2011-kd}; this model is typically only accurately known for already well-monitored volcanoes, for which the presented algorithms may be of more limited use.

				\subsubsection{Travel Time based Methods}\label{sec:travel_time_based_methods}
					Traveltime-based methods are the most common for locating seismic sources and have been widely used in volcano monitoring \citep{Lomax2001-xx,Saccorotti2007-ix,Woods2019-er}. The travel-time is the time taken for a seismic phase (e.g., P-wave, S-wave) to travel from the source to the receiver, and depends both on the locations of the source and receiver, and on the velocity model. If we assume a homogeneous velocity model with a velocity $v$ which represents the best estimate of the average velocity within the three dimensional region of interest, the observed arrival-time $t$ of the seismic waves can be calculated as:
					\begin{equation}
						t= t_\text{src} + \frac{1}{v} \left[
						(x_\text{rec} - x_\text{src})^2 + (y_\text{rec} - y_\text{src})^2 + (z_\text{rec} - z_\text{src})^2
						\right]^{\frac{1}{2}}
					\end{equation}
					where $t_\text{src}$ is the origin time, $(x_\text{src}, y_\text{src}, z_\text{src})$ is the source location, and $(x_\text{rec}, y_\text{rec}, z_\text{rec})$ is the receiver location. If we assume that there is little prior information on $t_\text{src}$ (the prior distribution is assumed to be uniform within certain spatial boundaries), the origin time can be integrated out since we assume a Gaussian data likelihood \citep{Tarantola1982-mc, Lomax2001-xx}. The model parameters are then only the source location $(x_\text{src}, y_\text{src}, z_\text{src})$, and arrival time observations become equivalent to travel time data. We treat the case of a heterogeneous velocity model seperately in section \ref{sec:heterogenous_velocity_model}, since it requires a more complex forward model.

					Uncertainty in the travel-time is expressed as the combination of errors in picking the arrival times of each seismic phase given the seismic waveform, and the contribution from the velocity model uncertainty which arises from the true seismic energy travelling through a heterogeneous medium which leads to a change in travel-time. This forward model error is constant in inverse problems but in experimental design problems it is unknown so we account for it through an uncertainty in the travel-time. A combined standard deviation of $\sigma_t$ can be calculated as:
					\begin{equation}
						\sigma_t(t)^2 	= \sigma_{\text{p}}^2 + t \cdot \sigma_\text{v}^2
					\end{equation}
					where $\sigma_{\text{p}}$ is the picking uncertainty and $\sigma_\text{v}$ is the relative velocity model uncertainty scaled by the travel-time $t$. Scaling the squared velocity model uncertainty by the travel-time is inspired by a Gaussian random walk model in which the mean squared distance (here travel-time noise) from the reference point (here the mean value) is proportional to the time.
					In a volcano setting the velocity model uncertainty term is typically much larger than the picking uncertainty term. To define it site specific information is necessary, but a value of 0.1 to 0.2 is a reasonable starting point as it corresponds to an average relative velocity model standard deviation of 10\% to 20\% from the specified homogeneous velocity model. This would result in a standard deviation of 0.1~s to 0.2~s for a travel-time of 1~s. Tests show that while the EIG and derived quantities are sensitive to the value of $\sigma_v$, the resulting optimal design is usually not particularly sensitive to it.
					
					Column one of Figure \ref{fig:likelihood_examples} shows the resulting likelihood for travel-time data from a synthetic true event for up to four stations. For a single station, the uncertainty spans a circle of equal travel-time. For multiple stations, the likelihood is the product of the likelihoods of the individual stations included in the design.

					Due to their emergent onset and lack of clear phase arrivals, long-period and tremor signals can not be located using travel-times based on phase arrival picks. However, several other methods, such as coherence based methods \citep{Ohminato1998-kg, Kawakatsu2000-pl, Dawson2004-qc, Almendros2003-ya, Saccorotti2021-cu} and back-propagation methods \citep{Kao2004-ar, Langet2014-mm} implicitly use travel-time information to locate sources. Therefore, optimising a network for travel-time observations also provides an optimal network for those methods.

				\subsubsection{Amplitude based Methods}\label{sec:amplitude_based_methods}

					Another common method for locating volcano-induced seismicity is amplitude source location (ASL), in which an attenuation model is assumed for a single wave type (surface or body waves) propagating isotropically in a homogeneous medium \citep{Saccorotti2021-cu, Battaglia2003-qp, Taisne2011-eu, Ogiso2012-ew, Kumagai2013-op, Kumagai2011-ak, Kumagai2009-pk, Ogiso2015-ym, Yamasato1997-yr, Jolly2002-ub, Caudron2018-mu, Carbone2008-jl}. The source location and strength can then be estimated by comparing observed amplitudes with those predicted from the attenuation model. If we ignore corrections for measurement site amplification, the observed amplitude $A$ decays as \citep{Kumagai2013-op, Morioka2017-ua}:
					\begin{equation}
						A = \frac{1}{r} \exp\left(-C \cdot t \right)
					\end{equation}
					where $r$ is the length of the ray path, and $C=\pi \frac{f}{Q}$ is the attenuation coefficient which depends on the frequency $f$, the quality factor $Q$, and the travel-time $t$ of the seismic phase. The length of the ray path $r$ as well as the travel-time $t$ depend on the source location $(x_\text{src}, y_\text{src}, z_\text{src})$ and receiver location $(x_\text{rec}, y_\text{rec}, z_\text{rec})$.

					We model the uncertainty in the amplitude through linearised error propagation \citep{Ku1966-us} of the uncertainty terms $\sigma_r$, $\sigma_t$, and $\sigma_Q$ for the ray path, travel-time and quality factor, respectively. The final uncertainty in the amplitude $\sigma_A$ is typically dominated by uncertainty in $\sigma_Q$. Internally, the final amplitude measurements and uncertainties are converted to log-space, which makes the Normal approximation in the $\mathrm{D}_N$ method more accurate.

					Column two in Figure \ref{fig:likelihood_examples} shows the resulting likelihood for amplitude data from a synthetic true event for up to four stations. While at first glance, this looks similar to the travel-time likelihood, the uncertainty is asymmetric in the radial direction due to the log-space conversion. For multiple stations the likelihood is again the product of the likelihoods of the individual stations.

				\subsubsection{Array Methods}\label{sec:array_methods}
					For the purposes of this paper we define an array as seismic array as a set of seismometers with inter-station spacings smaller than the wavelengths of interest, which typically have an aperture (largest distance between two stations) significantly smaller than the source-to-array distance \citep{Saccorotti2021-cu}. This is done in order to discriminate them from a network of seismometers that are processed independently of each other. For this study, we treat an array as	 an instrument at a single location which can measure a back azimuth $\theta$ and an incident angle $i$ (or alternatively, the two components $p_x$ and $p_y$ of the horizontal slowness vector). The back azimuth is the angle between the propagation direction of the arriving wavefront and the north direction, while the incident angle is the angle between propagation direction and the vertical direction. These quantities can be estimated straightforwardly using the angles between the source and the array for homogeneous models (or the gradient of the local travel-time grid for heterogeneous velocity models). In practice, the back azimuth and incident angle can be calculated using beamforming \citep[e.g.][]{Rost2002-mv, Leva2022-fv}, MUSIC \citep[e.g.][]{Inza2011-lu}, or other methods \citep[e.g.][]{Metaxian2002-sh,Di-Lieto2007-zl}.

					The uncertainty of the array measurements can either be defined through the standard deviation of the two angular measurements $\sigma_\theta$ and $\sigma_i$ or through the standard deviation of the two components of the slowness vector $\sigma_{p_x}$ and $\sigma_{p_y}$. The latter implies that there is an intuitive way to convert an array function to an uncertainty estimate. Within the code package, the data used are $p_x$ and $p_y$, so $\sigma_\theta$ and $\sigma_i$ are converted to $\sigma_{p_x}$ and $\sigma_{p_y}$ using linearised error propagation.

					Column three of Figure \ref{fig:likelihood_examples} shows the resulting likelihood for array data from a synthetic true event for up to four arrays. It is clear that this data type imposes substantially different constraints compared to travel-time and amplitude data: instead of a likelihood focused around a circle of equal travel-time or amplitude, we instead obtain a cone representing the uncertainty in the arrival direction of the seismic energy. And as before, the likelihood from multiple arrays is the product of the likelihoods of the individual arrays. Since the incident angle is very sensitive to the local seismic velocity and can change considerably for a gradient/heterogenous velocity model, it is also possible to use the back azimuth only. In this case, only $\sigma_\theta$ is used as the uncertainty in the array measurements.

				\subsubsection{Hybrid Data Types}\label{sec:hybrid_data_types}
					The fourth column in Figure \ref{fig:likelihood_examples} shows how likelihoods of the different data types can be combined. As above, these are simply point-by-point multiplications of the likelihoods of each sensor.

				\subsubsection{Important Assumptions}
					While the above data types capture the usual data used in volcano seismic monitoring, several important assumptions are made to make the design problem tractable, or because information necessary to model the problem more accurately is typically unavailable at the design stage. The most important assumptions are:

					\begin{itemize}
						\item Unless we have a subsurface velocity model available and use a considerably more expensive ray tracing method, we assume a homogeneous velocity model. While this is a severe assumption, it is often the only feasible way to simulate observed data in a volcano monitoring scenario. All of the experimental design methods can be extended to use a heterogenous velocity model, but this requires a more complex forward function to be available.
						\item We assume a Gaussian data likelihood. This is a common assumption in Bayesian inference and design studies, but both the choice of variances, and deviations from a Gaussian distribution can impact the quality of results \citep{Lomax2014-wr}. The experimental design methods could readily be be extended to use a different likelihood, but this would require much scenario-specific information that is hard to obtain, and which is almost never employed in practise even after data have been collected. Optimised designs are less important in already well-monitored scenarios, where the likelihood might be able to be modelled more accurately since there is usually a diminishing return for a larger number of stations/arrays.
						\item We assume that the uncertainty in each datum is independent of the others. While this is also a common assumption in Bayesian inference, it can significantly impact the results, especially since different data types at similar locations depend in some way on the same ray paths. The experimental design methods could readily be extended to include correlations between the data types, but estimating the correlations \textit{a priori} is usually difficult and, again, requires a lot of scenario-specific information. In practice, the impact of this assumption may be limited since experimental design algorithms typically spread out the locations of stations over the area of interest.
					\end{itemize}

					While these assumptions may seem restrictive, they are chosen to be practically implementable, andto deviate as little as possible from both the natural system and the assumptions typically made in Bayesian inference once data have been collected. In experimental design problems, we average over many samples of the prior distribution, which makes the impact of these assumptions less severe than in a typical inference problem.

		\section{Example Design Process}\label{sec:example_design_process}
			Finding an optimal experimental design depends on the specific scenario and the goals of the monitoring campaign. Instead of providing a one-size-fits-all solution, we provide a step-by-step guide to design an optimal monitoring network for any specific scenario. Nearly every step in the following example can be adjusted to specific needs. We provide a Jupyter notebook that guides practitioners through the process and serves as a template for each scenario. For more information see the data availability statement (section \ref{sec:data_availability}).
			\begin{figure*}[ht!]
				\centering
				\includegraphics[width=\textwidth]{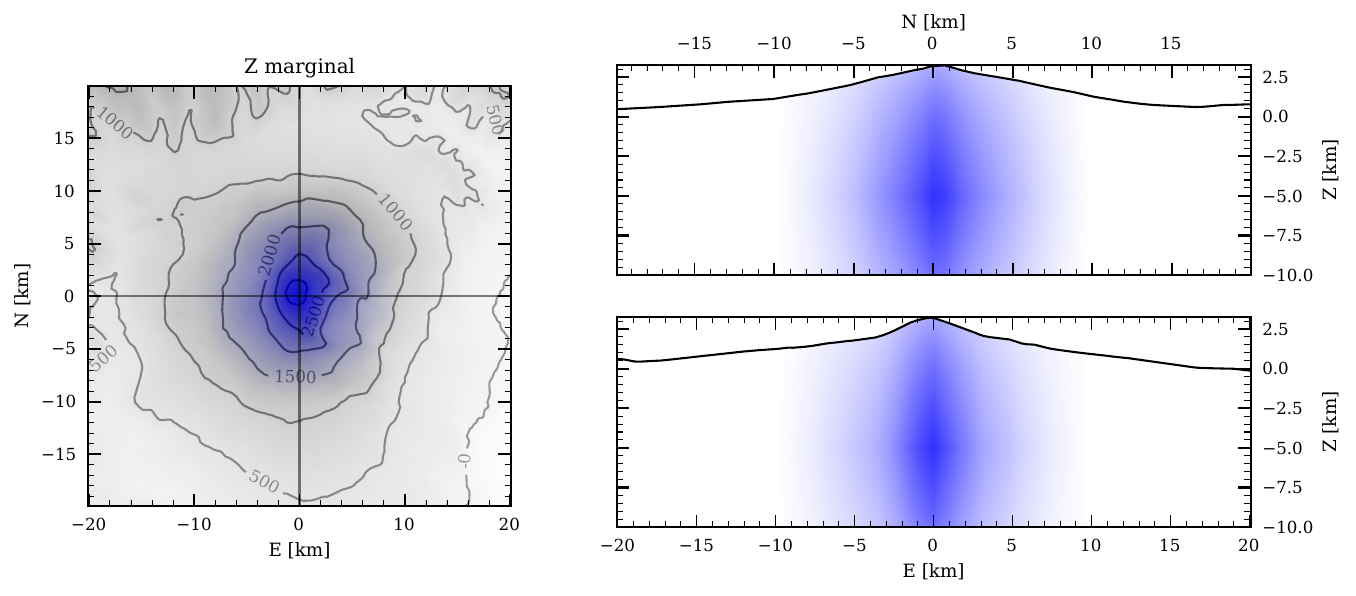}
				\caption{
					The prior distribution of the source locations for Etna. Dark blue indicates a high likelihood of seismicity, while light blue indicates a low likelihood. The elevation of the volcano is shown in the background. The left figure shows the marginal of the prior distribution along the depth axis, while the right figures show slices of the prior distribution at the Easting and Northing of the synthetic true event.
				}
				\label{fig:prior_model}
			\end{figure*}

			\subsubsection{Volcano Data}\label{sec:volcano_data}
				The Smithsonian Institution's Global Volcanism Program (GVP) \citep{Venzke2024-oa} provides a comprehensive database of volcanic activity, including eruption histories, reports, and data. We use this to obtain basic information about any volcano by providing its name, such as its location, type, and eruption history. This information is used for several of the following steps.
				The volcano used in this example is Etna, a stratovolcano in Italy. By defining a bounding box around the volcano's location, we use the OpenTopography API to obtain a digital elevation model (DEM) of the volcano. In this example, we use the SRTM15Plus \citep{Tozer2019-br}, which has a sufficient resolution for most volcano monitoring scenarios and unlike many other DEM's it includes bathymetric data. It is also possible to use a custom DEM, but this is often unnecessary.

			\subsubsection{Prior Information}\label{sec:prior_information}
				We can define a prior distribution for source locations based on the topographic information. We discretise the subsurface into a grid of cells, and a uniform prior distribution is assumed within each cell. For this example, we assume that the prior distribution is a Gaussian distribution with a standard deviation of 5~km in the Easting and Northing directions, and 8~km in the depth direction, centred the location of the volcano at a depth of 2~km. In addition, we assume that the prior distribution in each vertical column is proportional to the elevation at that point, which gives more weight to areas in which the volcano has a higher relative surface elevation. The result is, of course, a very simplified prior distribution, but it is easy to define, and since it is defined on a grid it is easy for any practitioner to adapt to a more complex distribution that takes a specific geological setting and other available information into account. Figure \ref{fig:prior_model} shows the prior distribution used in this example.

			\subsubsection{Design Space}\label{sec:design_space}
\begin{figure}[ht!]
					\centering
					\includegraphics[width=8.6cm]{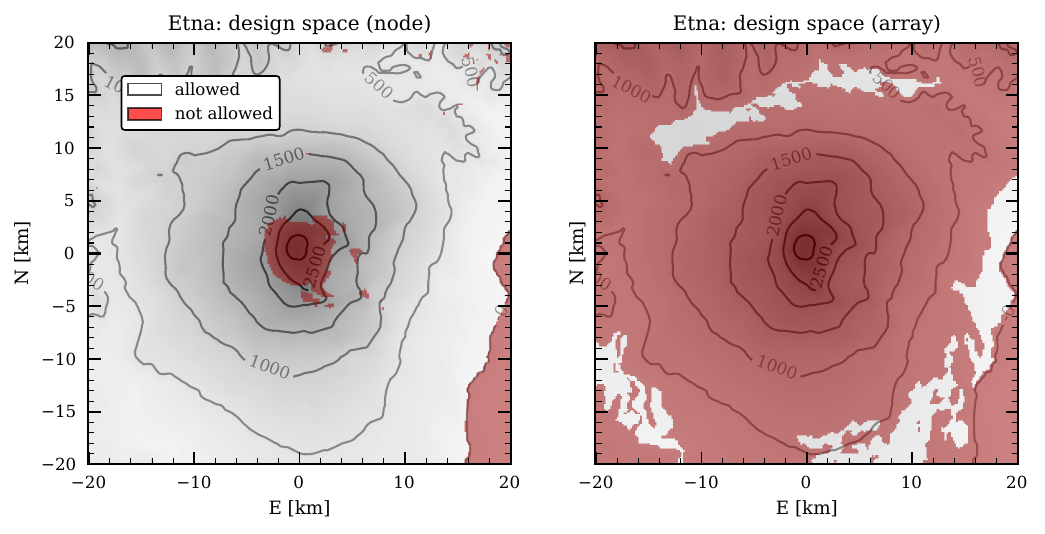}
					\caption{
						The design space for the nodal and array stations for the example volcano Etna. Red indicates that a station can not be placed there. The elevation of the volcano is shown by the background grey scale and the grey contour lines.
					}
					\label{fig:design_space}
				\end{figure}

				When optimising for the best design, we must consider numerous constraints present in each real-world scenario. In our code package, the design space (the set of all locations where a station or array can be placed) is defined on the grid of the (interpolated) DEM, where each cell is assigned a true or false value depending on whether or not a station can be placed there. In this example, we assume that nodal stations can only be placed in areas with an incline of less than 20 degrees, that array stations can be placed in regions with an incline of less than 3 degrees, that no station can be placed below sea level, that there is a safety margin of 3~km around the centre of the volcano if it erupted within the last ten years, and that arrays are placed in areas of at least 10~km$^2$ of flat ($<3$ degrees) area. Figure \ref{fig:design_space} shows the design spaces for the nodal and array stations, which implement the above constraints.

				Having the design space defined by general rules makes it easier to adapt to more complex scenarios with more constraints. Examples might include the use of satellite imagery to avoid recent lava flows, heavily wooded areas, urban areas, areas that are likely to experience high seismic noise, or any other area that is unsuitable for a seismic station. Using a genetic algorithm in the design optimisation process removes the need for constraints to have specific analytical properties, and so also makes it straightforward to include dynamic constraints such as the walking distance between stations, monetary constraints, or any other constraint that can be expressed as a function of the station locations.

			\subsubsection{Forward Function and Data Likelihood}\label{sec:forward_function}
				To construct the forward function, we need to define several parameters that govern the propagation of seismic waves in the subsurface, with often limited information available. For travel-time data we set the P-wave velocity to 3.5km/s \citep{Patane2002-on} and the corresponding uncertainties $\sigma_t$ to 0.01s and $\sigma_v$ to 0.1 (corresponds to a characteristic standard deviation of the velocity of $10\%$ along the ray path). For amplitude measurments, the quality factor $Q$ is set to 50, the frequency of interest $f$ to 2.0Hz, and the S-wave velocity is set to $v_s = v_p/\sqrt{3}$ \citep{Patane2002-on,Morioka2017-ua,Kumagai2009-pk}. A literature review, often including other, analogue volcanoes or geologies, is often necessary in order to estimate a suitable quality factor $Q$ and uncertainty for amplitude data. The low computational cost of the $\mathrm{D}_N$ method allows possible values to be tested rapidly, and in our experience, many different combinations of values typically result in similar experimental designs. The standard deviation of the quality factor is set to 10 in this example, and this, together with the uncertainties of the S-wave travel-time and the distance between the source and receiver allows the uncertainty in amplitude to be calculated. The uncertainty in the azimuth is set to 6 degrees \citep{Inza2011-lu}, and the incident angle is not used in this example. The back azimuth measurements require the least prior information. Within the code, the orientation of the array is corrected for the local topography.

			\subsubsection{Design Optimisation}\label{sec:design_optimisation}
				\begin{figure}[ht!]
					\includegraphics[width=8.6cm]{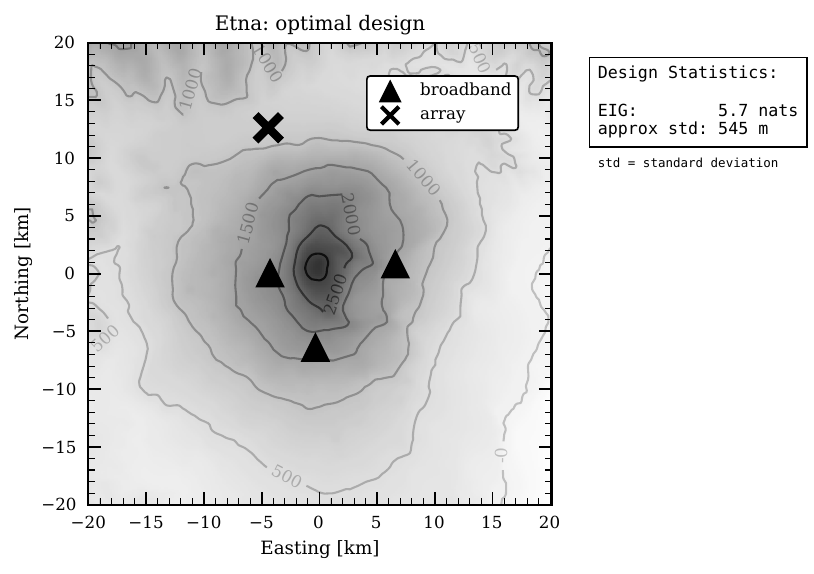}
					\caption{
						The optimal design for a network of three nodal stations and one array on Etna. The elevation of the volcano is shown in the background grey scale and the grey contour lines. For more information about the design statistics given in the box top-right, see section \ref{sec:design_appreciation}.
					}
					\label{fig:optimal_design}
				\end{figure}

				To find the optimal design, we use a genetic algorithm \citep{Holland1975-wf, Sambridge1993-sr, Gad2023-mq} and the EIG to be optimised is calculated using the $\mathrm{D}_N$ method. In this example, we optimise a network of three nodal stations (travel-time and amplitude information) and one array (travel-time, amplitude, and back azimuth information). Using a population size of 64 designs, around 200 generations (iterations of the algorithm) are needed for the optimisation to converge to a solution, with only slight improvements occurring thereafter up to around 600 generations. Figure \ref{fig:optimal_design} shows the optimal design for this scenario.

				On an average laptop, results are available within a few minutes for this example, if the $\mathrm{D}_N$ method is used; the NMC method would take tens of minutes for the same number of samples, or several days if the number of samples is increased by an order of magnitude which would be necessary for an unbiased result. Note that the forward function used here, while not fully optimised, can still be calculated extremely rapidly. The difference between the true EIG and the estimates from either the $\mathrm{D} _N $ or NMC method (for the number of samples used) is mostly constant for the seismic source location example, resulting in similarly well-performing designs \cite{Strutz2023-pl}. Therefore, both methods can be used in the design optimisation process.
				
				In some scenarios (e.g., prior distribution is far from Gaussian; a multimodal prior distribution; low noise levels), care must be taken when the $\mathrm{D}_N$ method is used so as not to violate the assumptions of a roughly Gaussian evidence too much. While this could be tested by plotting the evidence for a few designs, it is often easier in practice to use the NMC method and run another optimisation (with fewer generations) using the NMC method where the starting population is the optimal design found using the $\mathrm{D}_N$ method (this is implemented in the code package) to see whether the optimal design changes significantly. If it does, the $\mathrm{D}_N$ method may not be suitable for the scenario in question. Special care must be taken in the case of multimodal prior distributions (see \citet	{Strutz2023-pl} for an example of the $\mathrm{D}_N$ method failing in such a seismic source location scenario). Nevertheless, in general, the $\mathrm{D}_N$ method has proven to be robust in previous design optimisation studies \citep{Strutz2023-pl, Coles2013-vw, Bloem2020-gp}.

			\subsubsection{Design Analysis}\label{sec:design_appreciation}
\begin{figure*}[ht!]
					\centering
					\includegraphics[width=\textwidth]{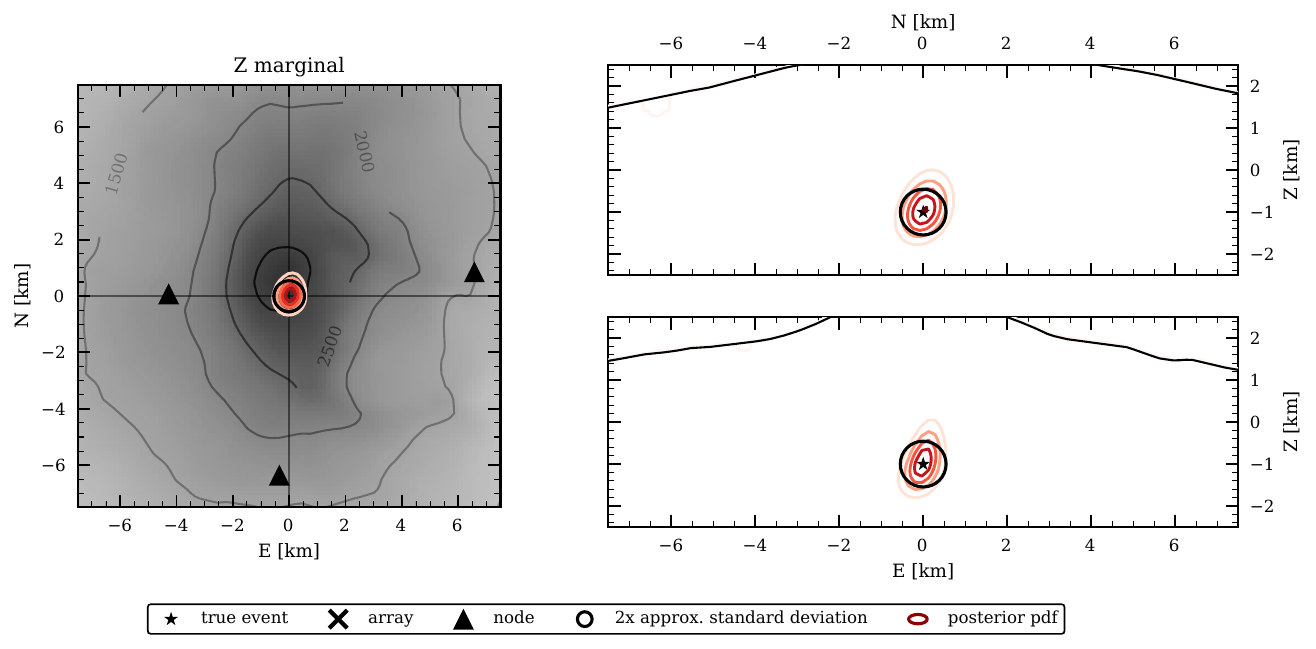}
					\caption{
						The posterior distribution for a synthetic true event for Etna, recorded with the optimal design. The synthetic true event is shown as a black star, and the uncertainty contours of the Bayesian posterior distribution are shown in red (high) to orange (low probability density). The expected standard deviation of the posterior distribution derived from the EIG is shown as a black circle with radius $\bar{\sigma}_\text{post}$. The left figure shows the marginal of the posterior distribution integrated over the depth axis. The right figures show slices of the posterior distribution at the Easting and Northing of the synthetic true event.
					}
					\label{fig:posterior_model}
				\end{figure*}

				After an optimal design has been computed it can be deployed. However, in many cases we would first like to analyse its performance in more detail, and possibly refine it either semi-automatically or manually. To understand the expected uncertainties in the inversion results using the final design, we can calculate the posterior distribution for an example event. An example result is shown in Figure \ref{fig:posterior_model}, where we observe that the (synthetic) true event lies well within the area of high posterior probability. By running several such tests, and by varying the design slightly, we can test how the uncertainty contours depend on the experimental design locations. As expected, the uncertainty in the vertical direction is substantially larger than in the horizontal directions since data is only collected on the surface at substantial offsets from the event. A receiver placed more directly above the event in question would conform more closely to the optimal design for a single event and allow more accurate estimates of the event depth \citep{Bloem2020-gp,Rabinowitz1990-cw, Steinberg1995-ox}, but in this case those locations are precluded (Figure \ref{fig:design_space}).

				The main benefit of the EIG calculation is that it provides a measure of the information gain expected over all possible true source locations (according to the prior distribution) instead of just one or a small subset of events. While we could use the EIG directly to compare different designs, it is an unintuitive quantity which is hard to interpret. If we assume that the posterior distribution is a multivariate isotropic Gaussian distribution (same standard deviation $\sigma_\text{post}$ for all parameters, and no inter-parameter correlations - see black circles overlying contours in Figure \ref{fig:posterior_model}), we can use the expected posterior information $ \bar{\mo{I}}_\text{post} = \mathbb{E}_{p(\bm{m})}\left[ \mo{I}_\text{post} \right] = \mo{EIG} - \mo{I}_\text{prior}$, to estimate the expected standard deviation 
						\begin{equation}
				\bar{\sigma}_\text{post}^2 = \exp{
				\frac{- \bar{\mo{I}}_\text{post}}{3} - \frac{1}{2} \left( 1 + \log(2 \pi) \right)}
						\end{equation}
				by using the analytic expression for the information in a multivariate Gaussian in Equation \ref{eqn:info_gaussian}. This expected standard deviation is an intuitive quantity that can be used to compare different designs and to analyse the expected uncertainty in the results. The effect of the isotropic assumption is that $\bar{\sigma}_\text{post}$ will often overestimate horizontal uncertainties and underestimate vertical uncertainties.

				We provide interactive plotting tools in the Jupyter notebook that accompanies this paper (see data availability statement in section \ref{sec:data_availability}), which allow the effects of different design locations on the posterior distribution and the expected standard deviation to be explored in real time. This makes it easy to refine the design and to include additional design constraints that are hard or impossible to include in the design space or the optimisation process.
				\begin{figure}[ht!]
					\centering
					\includegraphics[width=8.6cm]{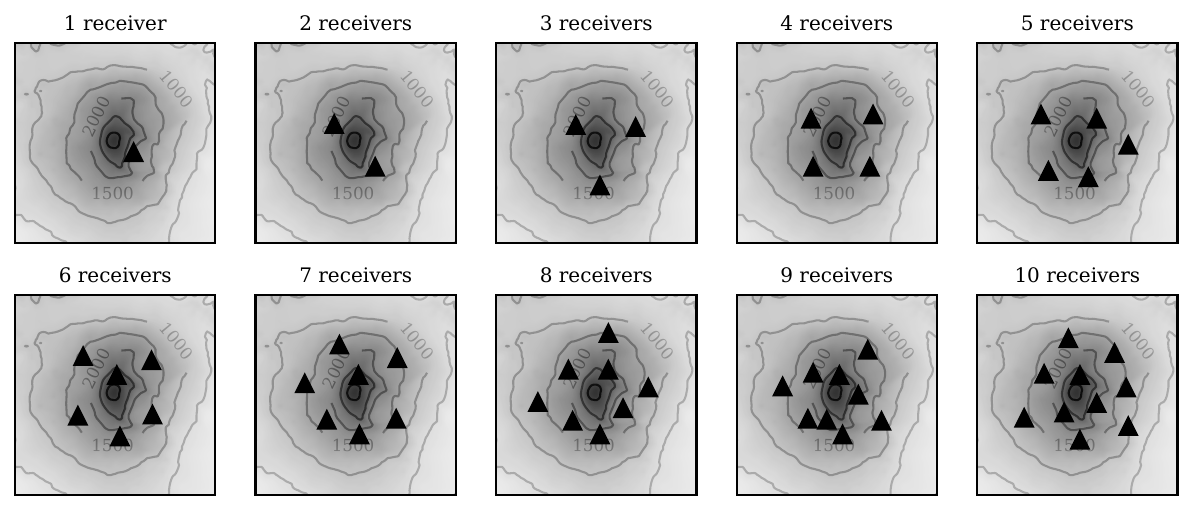}
					\caption{
						Example of optimal designs for Etna with up to 10 receivers. The coordinate axe span -15 to 15~km for both Easting and Northing.
					}
					\label{fig:example_designs}
				\end{figure}

\begin{figure}[ht!]
					\centering
					\includegraphics[width=8.6cm]{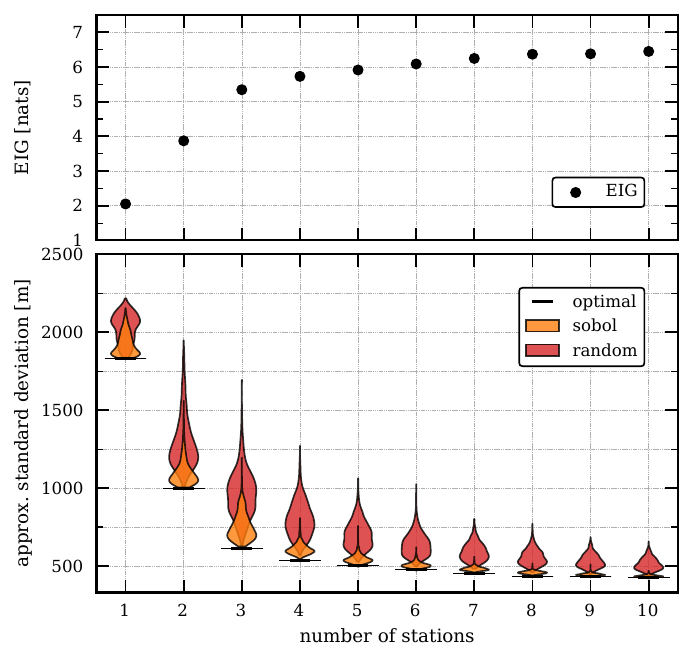}
					\caption{
						The expected standard deviation and the EIG of the posterior distribution as a function of the number of receivers deployed in optimal configurations for the example volcano, Etna (Fig. \ref{fig:example_designs}). The violin plots for random and sobol designs show the density of the expected standard deviation for 1000 random and sobol designs. The wider the violin plot, the more designs have that expected standard deviation. Optimal designs provide the lowest expected standard deviation in all cases.
						}
					\label{fig:uncertainty_goal}
				\end{figure}

				Another way to analyse the optimal design process is to observe how the expected information gain and standard deviation evolve as a function of receiver count. Optimal networks of up to 10 receivers which record travel-time and amplitude data, given the same parameters as in the previous example, are shown in Figure \ref{fig:example_designs}. We can calculate EIG and $\hat{\sigma}_\text{post}$ for each of these networks and plot them as a function of the number of receivers (Figure \ref{fig:uncertainty_goal}). This shows how many receivers are needed to reach any desired level of source location uncertainty.

				The $\hat{\sigma}_\text{post}$ results can also be used to compare the optimal designs to randomly placed, or approximately evenly spaced designs. To generate the random designs, we sample the design space according to a uniform distribution within the bounding box. The evenly spaced designs are generated by sampling a Sobol sequence - a quasi-random, space-filling sequence defined on $(0, 1)^D$ \citep{Sobol-1967-vb} which is then scaled by a factor sampled uniformly from 0 to 20~km. While the latter procedure is a complicated way to define reference space-filling designs, most of the resulting designs are closer to what we might consider to be a reasonable, uniformly distributed random design than are purely uniformly random designs, which is also reflected in the results. The expected standard deviation of 1000 of the random and Sobol designs are shown in Figure \ref{fig:uncertainty_goal} as violin plots. It is clear that the optimal designs are substantially better than the random designs, and that the Sobol designs are better than uniformly random designs, and indeed the latter approach the performance of optimal designs. However, for more than two receivers, an optimal design allows one to achieve the same uncertainty level as the mean of the Sobol designs with one receiver fewer, which is clearly desirable.
			
				For more than around eight receivers, the reduction in uncertainty for each additional receiver becomes relatively small. This property of diminishing returns is a common feature of optimal design problems \citep{Maurer1998-vd}, but our code package allows this effect to be quantified rapidly. Even if sufficient receivers are used to reach the point of diminishing returns (we refer to such cases as large-N designs, where N is the number of receivers), it is valuable to know how many receivers are required at a minimum to reach consistent performance for any approximately evenly spaced design. While the optimal large-N design process does not add much information in such cases, the confimation that N is indeed large enough is valuable information in itself.

		\section{Heterogenous Velocity Model}\label{sec:heterogenous_velocity_model}
\begin{figure}
				\centering
				\includegraphics[width=8.6cm]{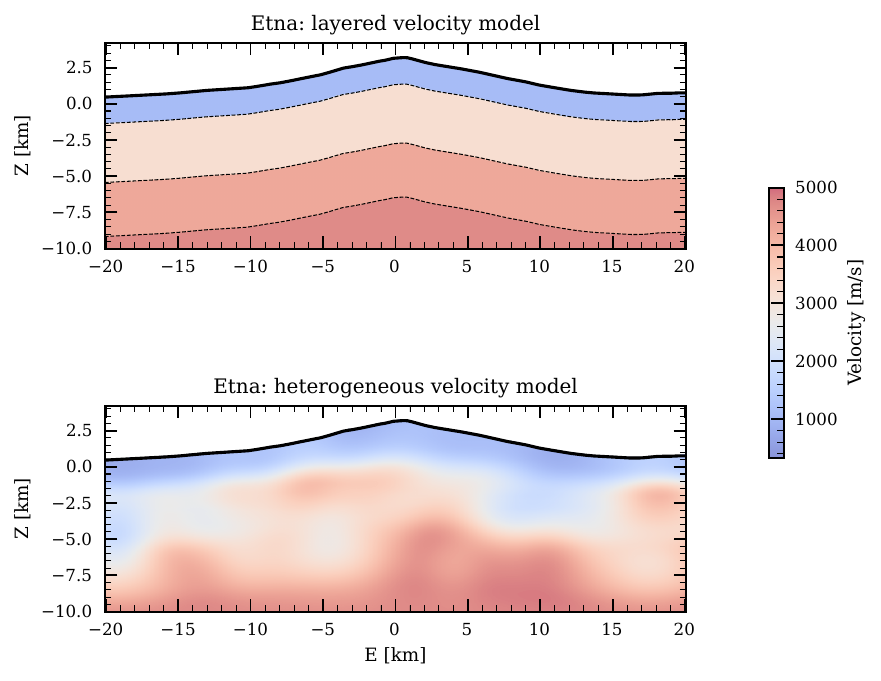}
				\caption{
					The layered and full 3D heterogeneous P-wave seismic velocity models used in this study. The layered model is the velocity model from \citet{Villasenor1998-mp}, and the heterogeneous model is a 3D model in which noise with a spatial wavelength of around 3~km and a relative amplitude of up to 50\% is added to a smoothed version of the layered model.
				}
				\label{fig:velocity_models}
			\end{figure}

			This section briefly explores how a heterogeneous seismic velocity model affects the formulation of the experimental design problem and the resulting optimal designs. The first immediate effect is that the forward function for all three data types is substantially more computationally expensive since an eikonal solver or seismic raytracer is needed to calculate travel-times, ray lengths and arrival angles. We use the open-source software package Pykonal \citep{White2020-nh} to calculate travel-time fields from which the other data types can be derived. To make the experimental design process more efficient, we precompute the data for every possible receiver/array location for a sufficient number of prior distribution samples such that results became stable under additional samples (typically around \ten{3} - \ten{4} samples) and store the results in a lookup table. We tested a layered model \citep{Villasenor1998-mp} and a full 3D heterogeneous velocity model in which we added variations with a spatial wavelength of around 3~km and a relative amplitude of up to 50\% to a smoothed version of the layered velocity model (none of the velocity models in the literature was openly available e.g.,\citet{Villasenor1998-mp, Patane2002-on}). Both models are shown in Figure \ref{fig:velocity_models}.

			Since the computation time for the data table lookups is in the order of the computation time for the NMC method, in this case we use the more accurate NMC method for the optimal design process. For this experiment, we only consider travel-time and array-derived data since calculating the large number of ray paths necessary for the amplitude measurements would be computationally expensive, and previous tests showed that this data had a small effect on the optimal design process. Again, we use 500 generations in the genetic algorithm with a population size of 64, which still enables us to calculate the travel-time table and the optimal design within a few hours on a standard laptop.
			\begin{figure}
				\centering
				\includegraphics[width=8.6cm]{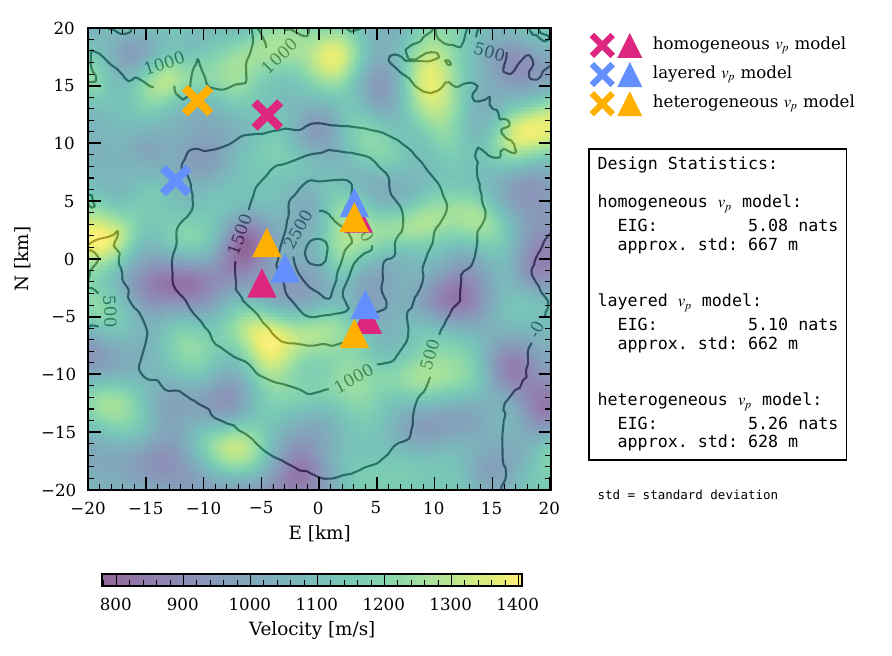}
				\caption{
					The optimal design for a homogeneous, layered and heterogeneous velocity model for three receivers and a single array. The EIG and approximated expected standard deviation are calculated using the heterogeneous velocity model for all three optimal designs and shown in the box on the right. The seismic velocity model at the surface is shown in the background. Crosses denote arrays and triangles denote nodal stations.
				}
				\label{fig:hom_lay_het_comparison}
			\end{figure}

			The results of the optimal design process for a homogeneous, layered and heterogeneous velocity model for three receivers and a single array, derived with the same settings, are shown in Figure \ref{fig:hom_lay_het_comparison}. EIG and approximated expected standard deviation are calculated with the heterogeneous velocity model for all three optimal designs. The performance of the homogeneous and layered models are very similar, but as expected the optimal design for the heterogeneous model performs better. While the relative placement of the nodes follows a similar pattern, the arrays are placed in different locations. Another observation not shown in the summary statistic is that the three designs perform very similarly for events at shallow or medium depth, and the increase in performance of the heterogeneous models is primarily due to improved performance for deep events.

			The results show that the optimal design process is relatively robust towards the choice of velocity model, but a performance increase can be achieved by going beyond a 1D layered media. This also suggests that care must be taken for 3D velocity models with high uncertainty, since the optimal design process will be sensitive to the velocity model employed.

		\section{Discussion and Conclusion}\label{sec:discussion_and_conclusion}
			We have presented a novel combination of experimental design methods for the optimal design of seismic monitoring networks for volcano monitoring. The methods are based on Bayesian experimental design to take the full uncertainty of the inversion process into account. We have shown that the expected standard deviation of the posterior distribution is a valuable quantity to evaluate the design quality, and present ways to interpret how much value an optimal design can add to post-survey inversion results.

			In the pursuit of accessibility, several assumptions were made to allow for a straightforward use, even if limited information is available. Most assumptions can be relaxed if more information is available, as shown in the example of the heterogeneous velocity model. The methods are also flexible enough to include more complex constraints and data types, but since those are typically scenario-specific, we have not included them in this work. Extending the methods and code package to design networks that are optimised to constrain the full source solution including the moment tensor is a challenging problem that is not addressed in this work. The methods presented here can be extended for this purpose, but it is currently not clear how this can be done in such a computationally efficient way.
		
			The main advance of this work is the use of both fast ($D_N$) and accurate (NMC) methods to calculate the expected information gain, and setting up the codes with access to public databases such that they can be deployed rapidly for any volcano in the world. This approach allows practitioners to optimise designs within minutes and have interactive tools available to refine the results. We hope that this work will enable more researchers to use optimal design methods without needing to be experts in the field, and will allow both experts and non-experts alike to use optimal designs even when responding rapidly to evolving eruptive scenarios.

		\section{Acknowledgements}\label{sec:acknowledgements}
			The implementations of the algorithms in this work would not have been possible without extensive use of open-source software. Not all of them have been included in the respective sections to ease readability. All the code was written in Python \citep{Van_Rossum2011-ep}, and the libraries NumPy \citep{Harris2020-nt}, SciPy \citep{Virtanen2020-ek}, Matplotlib \citep{Hunter2007-jw}, PyGAD \citep{Gad2023-mq}, and xarray \citep{Hoyer2017-nw} have been used extensively.

			This project has received funding from the European Union’s Horizon 2020 research and innovation programme under the Marie Skłodowska-Curie grant agreement no. 955515—SPIN ITN (www. spin-itn.eu).
			
			This work is based on data services provided by the OpenTopography Facility with support from the National Science Foundation under NSF Award Numbers 1948997, 1948994 \& 1948857.

		\section{Data Availability}\label{sec:data_availability}
			All code necessary to reproduce the results is available at \url{https://github.com/dominik-strutz/WoWED-volcano}. The interactive Jupyter notebook that accompanies this paper and guides the user through the design process is available in the same repository. All other data is publicly available and referenced in the respective sections.

		\bibliography{bibliography.bib}

		\appendix
			\section{Shannon Information}\label{app:shannon_information}

				Shannon information \citep{Shannon1948-of} is an intuitive measure of information with several beneficial properties (\eg linear additivity of information from independent sources). The Shannon information $\mo{I}[\cdot]$ of an arbitrary continuous probability density function $p(x)$ is defined as
				\begin{align}
					\begin{split}
					\mo{I} \left[p(x) \right] &= \mathbb{E}_{p(x)} \left[ \log _{b}\left(p(x) \right) \right] \\
					&= \int_{\mathcal{X}} p(x) \log _{b}(p(x)) d x
					\end{split}
					\label{eqn:information}
				\end{align}
				where $x\in \mathcal{X}$ is a random variable distributed according to $p(x)$ and $\mathbb{E}_{p(x)}$ is the expectation with respect to $p(x)$, which is defined by the right-most expression. Depending on the context, information is also often expressed through the entropy $\mo{H}[p(x)] = - \mo{I} \left[p(x) \right]$, where entropy $\mo{H}$ is defined to be the negative of either expression on the right of Equation \eqref{eqn:information}. This absolute information measure can be extended to the relative information content of one pdf relative to another, also called the Kullback-Leibler (KL) divergence \citep{Kullback1951-pq}
				\begin{align}
					\text{KL}(P||Q) = \int_\mathcal{X} p(x) \log\left( \frac{p(x)}{q(x)} \right) d x
				\end{align}
				For further information on the properties of information, the reader is referred to \citet{Cover2006-bi}.

			\section{Rearanging the Expected Information Gain}\label{app:rearange_EIG}

				\begin{align}
					\mo{EIG}(\bm{\xi}) &= \mathbb{E}_{p(\bm{d} | \bm{\xi})} \left[\mo{I}[p(\bm{m} \, | \, \bm{d},\bm{\xi})] - \mo{I}[p({\bm{m}})] \right] \\
									&= \mathbb{E}_{p(\bm{d}, \bm{m} | \bm{\xi})} \left[ \log \frac{p(\bm{m} \, | \, \bm{d},\bm{\xi})}{p({\bm{m}})} \right]\\
									&= \mathbb{E}_{p(\bm{d}, \bm{m} | \bm{\xi})} \left[ \log \frac{p(\bm{m}, \bm{d} \, | \, \bm{\xi})}{p({\bm{m}}) p(\bm{d} \, | \,\bm{\xi})} \right]\\
									&= \mathbb{E}_{p(\bm{d}, \bm{m} | \bm{\xi})} \left[ \log \frac{p(\bm{d} \, | \, \bm{m},\bm{\xi})}{p(\bm{d} \, | \, \bm{\xi})} \right]\\
					\mo{EIG}(\bm{\xi}) &= \mathbb{E}_{p(\bm{m})}\left[ \operatorname{I}[p(\bm{d} \, | \, \bm{m},\bm{\xi})] - \mo{I}[p(\bm{d} \, | \,\bm{\xi})] \right]
				\end{align}

		\end{document}